# Big data, bigger dilemmas:
# A critical review

Hamid Ekbia[1], Michael Mattioli[2], Inna Kouper[1,3], G. Arave[1], Ali Ghazinejad[1], Timothy Bowman[1], Venkata Ratandeep Suri[1], Andrew Tsou[1], Scott Weingart[1], and Cassidy R. Sugimoto*[1]

{hekbia, mmattiol, inkouper, garave, alighazi, tdbowman, vsuri, atsou, sbweing, sugimoto}@indiana.edu

*corresponding author

1. Center for Research on Mediated Interaction (CROMI)
School of Informatics and Computing
Indiana University Bloomington

2. Center for Intellectual Property
Maurer School of Law
Indiana University Bloomington

3. Data to Insight Center
Pervasive Technology Institute
Indiana University

*Abstract*

The recent interest in Big Data has generated a broad range of new academic, corporate, and policy practices along with an evolving debate amongst its proponents, detractors, and skeptics. While the practices draw on a common set of tools, techniques, and technologies, most contributions to the debate come either from a particular disciplinary perspective or with an eye on a domain-specific issue. A close examination of these contributions reveals a set of common problematics that arise in various guises in different places. It also demonstrates the need for a critical synthesis of the conceptual and practical dilemmas surrounding Big Data. The purpose of this article is to provide such a synthesis by drawing on relevant writings in the sciences, humanities, policy, and trade literature. In bringing these diverse literatures together, we aim to shed light on the common underlying issues that concern and affect all of these areas. By contextualizing the phenomenon of Big Data within larger socio-economic developments, we also seek to provide a broader understanding of its drivers, barriers, and challenges. This approach allows us to identify attributes of Big Data that need to receive more attention — autonomy, opacity, and generativity, disparity, and futurity — leading to questions and ideas for moving beyond dilemmas.

## INTRODUCTION

"Big Data" has become a topic of discussion in various circles, arriving with a fervor that parallels some of the most significant movements in the history of computing — e.g., the development of personal computing in the 1970's, the World Wide Web in the 1990's, and social media in the 2000's. In academia, the number of dedicated venues (journals, workshops, or conferences), initiatives, and publications on this topic reveal a continuous and consistent growing trend.

Figure 1, which shows the number of scholarly, trade and mass media publications across five academic databases in the last six years with Big Data in their title or as a keyword, is illustrative of this trend.

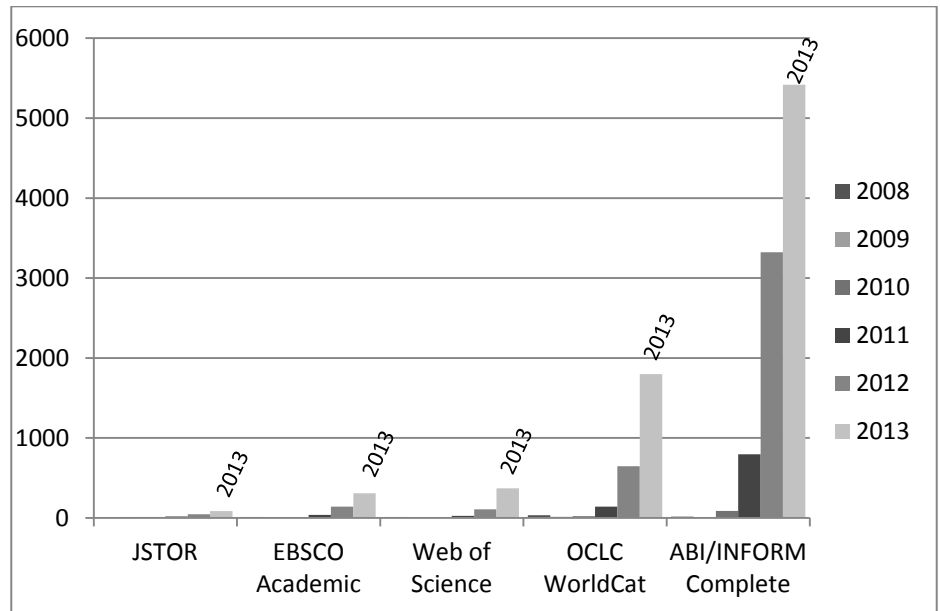

Figure 1. Number of publications with the phrase "big data" in five academic databases between 2008 and 2013.

Along with academic interest, practitioners in business, industry, and government have found that Big Data offers tremendous opportunities for commerce, innovation, and social engineering. The World Economic Forum (2011), for instance, dubbed data a new "asset class," and some commentators have mused that data is "the new oil" (e.g., Rotella, 2012). These developments provide the impetus for the creation of the required technological infrastructure, policy frameworks, and public debates in the U.S., Europe, and other parts of the world (National Science Foundation, 2012; OSP, 2012; Williford & Henry, 2012). The growing Big Data trend has also prompted a series of reflective engagements with the topic that, although few and far between, introduce stimulating insights and critical questions (e.g., Borgman, 2012; boyd & Crawford, 2012). The interest within the general public has also intensified due to the blurring of boundaries between data produced *by* humans and data *about* humans (Shilton, 2012).

These developments reveal the diversity of perspectives, interests, and expectations regarding Big Data. While some academics see an opportunity for a new area of study and even a new kind of "science," for instance, others emphasize novel methodological and epistemic approaches, and still others see potential risks and pitfalls. Industry practitioners, trying to make sense of their own practices, by contrast, inevitably draw on the available repertoire of computing processes and techniques—abstraction, decomposition, modularization, machine learning, worse-case analysis, etc. (Wing, 2006)—in dealing with the problem of what to do when one has "lots of data to store, or lots of data to analyze, or lots of machines to coordinate" (White, 2012, p. xvii). Finally, legal scholars, philosophers of science, and social commentators seek to inject a critical perspective by identifying the conceptual questions, ethical dilemmas, and socio-economic challenges brought about by Big Data.

These variegate accounts point to the need for a critical synthesis of perspectives, which we propose to analyze and explain as a set of dilemmas.[1] In our study of the diverse literatures and practices related to Big Data, we have observed a set of recurrent themes and problematics. Although they are often framed in terminologies that are discipline-specific, issue-oriented, or methodologically-informed, these themes ultimately speak to the same underlying frictions. We see these frictions as inherent to the modern way of life, embodied as they currently are in the socio-historical developments associated with digital technologies—hence, our framing them as *dilemmas* (Ekbia, Kallinikos, & Nardi, 2014). Some of these dilemmas are novel, arising from the unique phenomena and issues that have recently emerged due to the deepening penetration of the social fabric by digital information, but many of them are the reincarnation of some of the erstwhile questions and quandaries in scientific methodology, epistemology, aesthetics, ethics, political economy, and elsewhere. Because of this, we have chosen to organize our discussion of these dilemmas according to their origin, incorporating their specific disciplinary or technical manifestations as special cases or examples that help to illustrate the dilemmas on a *spectrum* of possibilities. Our understanding of dilemmas, as such, is not binary; rather, it is based on gradations of issues that arise with various nuances in conceptual, empirical, or technical work.

Before embarking on the discussion of dilemmas faced on these various dimensions, we must specify the scope of our inquiry into Big Data, the definition of which is itself a point of contention.

**CONCEPTUALIZING BIG DATA**

A preliminary examination of the debates, discussions, and writings on Big Data demonstrates a pronounced lack of consensus about the definition, scope, and character of what falls within the purview of Big Data. A meeting of the Organization for Economic Co-operation (OECD) in February of 2013, for instance, revealed that all 150 delegates had heard of the term "Big Data," but a mere 10% were prepared to offer a definition (Anonymous, 2013). Perhaps worryingly, these delegates "were government officials who will be called upon to devise policies on supporting or regulating Big Data" (Anonymous, 2013, para. 1).

A similar uncertainty is prevalent in academic writing, where writers refer to the topic as "a moving target" (Anonymous, 2008, p.1) that "can be 'big' in different ways" (Lynch, 2008, p. 28). Big Data has been described as "a relative term" (Minelli, Chambers, & Dhiraj, 2013, p. 9) that is intended to describe "the mad, inconceivable growth of computer performance and data storage" (Doctorow, 2008, p. 17). Such observations led Floridi (2012) to the conclusion that "it is unclear what exactly the term 'Big Data' means and hence refers to" (p. 435). To give shape to this mélange of definitions, we identify three main categories of definitions in the current literature: (i) Product-oriented with a quantitative focus on the size of data; (ii) Process-oriented with a focus on the processes involved in the collection, curation, and use of data; and (iii) Cognition-oriented with a focus on the way human beings, with their particular cognitive capacities, can relate to data. We review these perspectives, and then provide a fourth—that is, a social movement perspective—as an alternative conceptualization of Big Data.

---

[1] By synthesis, we intend the inclusion of various perspectives regardless of their institutional origin (academia, government, industry, media, etc.). By critical, we mean a kind of synthesis that takes into account historical context, philosophical presuppositions, possible economic and political interests, and the intended and unintended consequences of various perspectives. By dilemma, we mean a situation that presents itself as a set of indeterminate outcomes that do not easily lend themselves to a compromise or resolution.

**The Product-oriented Perspective: The Scale of Change**

The product-oriented perspective tends to emphasize the attributes of data, particularly its size, speed, and structure or composition. Typically putting the size in the range of petabytes to exabytes (Arnold, 2011; Kaisler, Armour, Espinosa, & Money, 2013; Wong, Shen, & Chen, 2013), or even yottabytes (a trillion terabytes) (Bollier, 2010, p. vii), this perspective derives its motivation from estimates of increasing data flows in social media and elsewhere. On a daily basis, Facebook collects more than 500 terabytes of data (Tam, 2012), Google processes about 24 petabytes (Davenport, Barth, & Bean, 2012, p. 43), and Twitter, a social media platform known for the brevity of its content, produces nearly 12 terabytes (Gobble, 2013, p. 64). Nevertheless, the threshold for what constitutes Big Data is still contested. According to Arnold (2008), "100,000 employees, several gigabytes of email per employee" does not qualify as Big Data, whereas the vast amounts of information processed on the Internet—e.g., "8,000 tweets a second," or Google "receiv[ing] 35 hours of digital video every minute"—*do* qualify (p. 28).

Another motivating theme from the product-oriented perspective is to adopt an historical outlook, comparing the quantity of data present now with what was available in the recent past. Ten years ago, a gigabyte of data seemed like a vast amount of information, but the digital content on the Internet is currently estimated to be vloseclose to five hundred billion gigabytes. On a broader scale, "it is estimated that humanity accumulated 180 exabytes of data between the invention of writing and 2006. Between 2006 and 2011, the total grew ten times and reached 1,600 exabytes. This figure is now expected to grow fourfold approximately every 3 years" (Floridi, 2012, p. 435). Perhaps most illustrative of the comparative trend is the growth of astronomical data:

> When the Sloan Digital Sky Survey started work in 2000, its telescope in New Mexico collected more data in its first few weeks than had been amassed in the entire history of astronomy. Now, a decade later, its archive contains a whopping 140 terabytes of information. A successor, the Large Synoptic Survey Telescope, due to come on stream in Chile in 2016, will acquire that quantity of data every five days. (Anonymous, 2010c, para. 1).

Some variations of the product-oriented view take a broader perspective, highlighting the characteristics of data that go beyond sheer size. Thus, Laney (2001), among others, advanced the now-conventionalized tripartite approach to Big Data—namely, that data changes along three dimensions: volume, velocity, and variety (others have added value and veracity to this list). The increasing breadth and depth of our logging capabilities lead to the increase in data volume, the increasing pace of interactions leads to increasing data velocity, and the multiplicity of formats, structures and semantics in data leads to increasing data variety. An NSF solicitation to advance Big Data techniques and technologies encompassed all three dimensions, and referred to Big Data as "large, diverse, complex, longitudinal, and/or distributed data sets generated from instruments, sensors, Internet transactions, email, video, click streams, and/or all other digital sources available today and in the future" (Core Techniques and Technologies, 2012). Similarly, Gobble (2013) observes that "bigness is not just about size. Data may be big because there's too much of it (volume), because it's moving too fast (velocity), or because it's not structured in a usable way (variety)" (p. 64) (see also Minelli, Chambers, & Dhiraj, 2013, p. 83).

In brief, the product-oriented perspective highlights the novelty of Big Data largely in terms of the attributes of the data themselves. In so doing, it brings to forth the scale and magnitude of the changes brought about by Big Data, and the consequent prospects and challenges generated by these transformations.

**The Process-Oriented Perspective: Pushing the Technological Frontier**

The process-oriented perspective underscores the novelty of processes that are involved, or are required, in dealing with Big Data. The processes of interest are typically computational in character, relating to the storage, management, aggregation, searching, and analysis of data, while the motivating themes derive their significance from the opacity of Big Data. Big Data, boyd and Crawford (2011) posit, "is notable not because of its size, but because of its relationality to other data" (p. 1). Wardrip-Fruin (2012) emphasizes that with data growth human-designed and human-interpretable computational processes increasingly shape our research, communication, commerce, art, and media. Wong, Shen, and Chen (2012) further argue that "when data grows to a certain size, patterns within the data tend to become white noise" (p. 204).

According to this perspective, it is in dealing with the issues of opacity, noise, and relationality that technology meets its greatest challenge and perhaps opportunity. The overwhelming amounts of data that are generated by sensors and logging systems such as cameras, smart phones, financial and commercial transaction systems, and social networks, are hard to move and protect from corruption or loss (Big Data, 2012; Kaisler, Armour, Espinosa, and Money, 2013). Accordingly, Kraska (2013) defines Big Data as "when the normal application of current technology doesn't enable users to obtain timely, cost-effective, and quality answers to data-driven questions" (p. 85), and Jacobs (2009) suggests that Big Data refers to "data whose size forces us to look beyond the tried-and-true methods that are prevalent at that time" (p. 44). Rather than focusing on the size of the output, therefore, these writers consider the processing and storing abilities of technologies associated with Big Data to be its defining characteristic.

In highlighting the challenges of processing Big Data, the process perspective also emphasizes the requisite technological infrastructure, particularly the technical tools and programming techniques involved in the creation of data (Vis, 2013), as well as the computational, statistical, and technical advances that need to be made in order to analyze it. Accordingly, current technologies are deemed insufficient for the proper analysis of Big Data, either because of economic concerns or the limitations of the machinery itself (Trelles, Prins, Snir, & Jansen, 2011). This is partly because Big Data encompasses both structured and unstructured data of all varieties: text, audio, video, click streams, log files, and more (White, 2011, p. 21). By this light, "the fundamental problems ... are rendering efficiency and data management … [and] to reduce the complexity and number of geometries used to display a scene as much as possible" (Pajarola, 1998, p. 1). Complexity reduction, as such, constitutes a key theme of the process-oriented perspective, driving the current interest in visualization techniques (Ma & Wang, 2009; see also the section on Aesthetic Dilemmas below).

To summarize, by focusing on those attributes of Big Data that derive from complexity-enhancing structural and relational considerations, the process-oriented perspective seeks to push the frontiers of computing technology in handling those complexities.

**The Cognition-Oriented Perspective: Whither the Human Mind?**

Finally, the cognition-based perspective draws our attention to the challenges that Big Data poses to human beings in terms of their cognitive capacities and limitations. On one hand, with improved storage and processing technologies, we become aware of how Big Data can help address problems on a macro-scale. On the other hand, even with theories of how a system or a phenomenon works, its interactions are so complex and massive that the human brains simply cannot comprehend them (Weinberger, 2012). Based on an analogy with the number of words a human being might hear in their lifetime—"surely less than a terabyte of text"—Horowitz (2008) refers to the concerns of a computer scientist: "Any more than that and it becomes incomprehensible by a single person, so we have to turn to other means of analysis: people working together, or computers, or both." Kraska (2013) invokes another analogy to make a similar point: "A [terabyte] of data is nothing for the US National Security Agency (NSA), but is probably a lot for an individual" (p. 85).

In this manner, the concern with the limited capacity of the human mind to make sense of large amounts of data turns into a thematic focus for the cognition-oriented perspective. In particular, the implications for the conduct of science have become a key concern to commentators such as Howe et al. (2008), who observe that "such data, produced at great effort and expense, are only as useful as researchers' ability to locate, integrate and access them" (p. 47). This attitude echoes earlier visions underlying the development of scholarly systems. Eugene Garfield, in his construction of what would become the Web of Science, stated that his main objective was the development of "an information system which is economical and which contributes significantly to the process of information discovery—that is, the correlation of scientific observations not obvious to the searchers" (Garfield, 1963, p. 201).

Others, however, see an opportunity for new modes of scholarly activity—for instance, engaging in distributed and coordinated (as opposed to collaborative) work (Waldrop, 2008). By this definition, the ability to distribute activity, store data, and search and cross-tabulate becomes a more important distinction in the epistemology of Big Data than size or computational processes alone (see the section on Epistemological Dilemmas below). boyd and Crawford (2012), for instance, argue that Big Data is about the "capacity to search, aggregate, and cross-reference large data sets"(p. 665), an objective that have been exemplified in large-scale observatory efforts in biology, ecology and geology (Aronova, Baker, and Oreskes, 2010; Lin, 2013). .

In short, the cognition-oriented perspective conceptualizes Big Data as something that exceeds human ability to comprehend and therefore requires mediation through trans-disciplinary work, technological infrastructures, statistical analyses, and visualization techniques to enhance interpretability.

**The Social Movement Perspective: The Gap between Vision and Reality**

The three perspectives that we have identified present distinct conceptualizations of Big Data, but they also manifest the various motivational themes, problematics, and agendas that drive each perspective. None of these perspectives by itself delineates the full scope of Big Data, nor are they mutually exclusive. However, in their totality, they provide useful insights and a good starting point for our inquiry. We notice, in particular, that little attention is paid in these perspectives to the socio-economic, cultural, and political shifts that underlie the phenomenon of Big Data, and that are, in turn, enabled by it. Focusing on these shifts allows for a different line of inquiry—one that explains the diffusion of technological innovation as a "computerization movement" (Kling & Iacono, 1995).

The development of computing seems to have followed a recurring pattern wherein an emerging technology is promoted by loosely organized coalitions that mobilize groups and organizations around a utopian vision of a preferred social order. This historical pattern is the focus of the perspective of "computerization movement," which discerns a parallel here with the broader notion of "social movement" (De la Porta & Diani, 2006). Rather than emphasizing the features of technology, organizations, and environment, this perspective considers technological change "in a broader context of interacting organizations and institutions that shape utopian visions of what technology can do and how it can be used" (Elliot & Kraemer, 2008, p. 3). A prominent example of computerization movements is the nationwide mobilization of Internet access during the Clinton-Gore administration in the U.S. around the vision of a world where people can live and work at the location of their choice (The White House, 1993). The gap between these visions and socio-economic and cultural realities, along with the political strife associated with the gap, is often lost in articulations of these visions.

A similar dynamic is discernible in the case of Big Data, giving support to a construal of this phenomenon as a computerization movement. This is most evident in the strategic alliances formed around Big Data amongst heterogeneous players in business, academia, and government. The complex eco-system in which Big Data technologies are developed is characterized by a symbiotic relationship between technology companies, the open source community, and universities (Capek, Frank, Gerdt, & Shields, 2005). *Apache Hadoop,* the widely adopted open source framework in Big Data research, for instance, was developed based on contributions from the open source community and IT companies such as *Yahoo* using schemes such as the *Google File System* and *MapReduce*—a programming model and an associated implementation for processing and generating large data sets (Harris, 2013).

Another strategic partnership was shaped between companies, universities, and federal agencies in the area of education, aimed at preparing the next generation of Big Data scientists. In 2007, Google, in collaboration with IBM and the U.S. National Science Foundation (in addition to participation from major universities) began the *Academic Cluster Computing Initiative,* offering hardware, software, support services, and start-up grants aimed at improving student knowledge of highly parallel computing practices and providing the skills and training necessary for the emerging paradigm of large-scale distributed computing (Ackerman, 2007; Harris, 2013). In a similar move, IBM, as part of its *Academic Initiative,* has fostered collaborative partnerships with a number of universities, both in the U.S. and abroad, to offer courses, faculty awards, and program support focusing on Big Data and analytics (Gardner, 2013; International Business Machines [IBM], 2013). As of 2013, nearly 26 universities now offer courses related to Big Data science across the U.S.; these courses, accordingly, are supported by a number of IT companies (Cain-Miller, 2013).

The driving vision behind these initiatives draws parallels between them and "the dramatic advances in supercomputing and the creation of the Internet," emphasizing "our ability to use Big Data for scientific discovery, environmental and biomedical research, education, and national security" (OSP, 2012, para. 3). Business gurus, management professionals, and many academics seem to concur with this vision, seeing the potential in Big Data to quantify and thereby change various aspects of contemporary life (Mayer-Schonberger & Cukier, 2013), "to revolutionize the art of management" (Ignatius, 2012), or to be part of a major transformation that requires national effort (Riding the wave, 2010). As with previous computerization movements, however, there are significant gaps between these visions and the realities on the ground.

As suggested earlier, we propose to explore these gaps as a set of dilemmas on various dimensions, starting with the most abstract philosophical and methodological issues, and moving through specifically technical matters toward legal, ethical, economical, and political questions. These dilemmas, as we shall see below, are concerned with foundational questions about what we know about the world and about ourselves (epistemology), how we gain that knowledge (methodology) and how we present it to our audiences (aesthetics), what techniques and technologies are developed for these purposes (technology), how these impact the nature of privacy (ethics) and intellectual property (law), and what all of this implies in terms of social equity and political control (political economy).

**EPISTEMOLOGICAL DILEMMAS**

The relationship between data—what the world presents to us through our senses and instruments—and our knowledge of the world—how we understand and interpret what we get—has been a philosophically tortuous one with a long history. The epistemological issues introduced by Big Data should be philosophically contextualized with respect to this history.

According to some philosophers, the distinction between data and knowledge has largely to do with the non-evident character of "matters of fact," which can only be understood through the relation of cause and effect (unlike abstract truths of geometry or arithmetic). For David Hume, for example, this latter relation "is not, in any instance, attained by reasonings *a priori*; but arises entirely from experience, when we find that any particular objects are constantly joined with each other" (Hume, 1993/1748, p. 17). Hume's proposition is meant to be a corrective response to the rationalist trust in the power of human reason—e.g., Descartes' (1913) dictum "that, touching the things which our senses do not perceive, it is sufficient to explain how they can be" (p. 210). In making this statement, Descartes points to the gap between the "reality" of the material world and the way it appears to us—e.g., the perceptual illusion of a straight stick that *appears* as bent when immersed in water, or of the retrograde motion of the planet Mercury that *appears* to reverse its orbital direction. The solution that he proposes, as with many other pioneers of modern science (Galileo, Boyle, Newton, and others), is for science to focus on "primary qualities" such as extension, shape, and motion, and to forego "secondary qualities" such as color, sound, taste, and temperature (van Fraassen, 2008). Descartes' solution was at odds, however, with what Aristotelian physicists adhered to for many centuries—namely, that science must explain how things happen by demonstrating that they *must* happen in the way they do. The solution undermined, in other words, the criterion of universal necessity that was the linchpin of Aristotelian science.

This "flaunting of previously upheld criteria" (van Fraassen, 2008, p. 277) has been a recurring pattern in the history of science, which is punctuated by episodes in which earlier success criteria for the completeness of scientific knowledge are rejected by a new wave of scientists and philosophers who declare victory for their own ways according to some new criteria. This, in turn, provokes complaints and concerns by the defendants of the old order, who typically see the change as a betrayal of some universal principle: Aristotelians considered Descartes' mechanics as violating the principle of necessity, and Cartesians blamed Newton's physics and the idea of action at a distance as a violation of the principle of determinism, as did Newtonians who derided Quantum Mechanics for that same reason. Meanwhile, the defendants of non-determinism, seeking new success criteria, introduced the *Common Cause Principle*, which gave rise to a new criterion of success that van Fraassen (2008) calls the *Appearance-from-Reality* criterion (p. 280). This criterion is based on a key distinction between *phenomenon* and *appearance* —i.e., between "observable entities (objects, events, processes, etc.) of any sort [and] the contents of measurement outcomes" (van Fraassen, 2008, p. 283). Planetary motions, for instance, constitute a phenomenon that *appears* differently to us terrestrial observers depending on our measurement techniques and technologies—hence, the difference between Mercury's unidirectional orbit and its appearance of retrograde motion. Another example is the *phenomenon* of micro-motives (e.g., individual preferences in terms of neighborhood relations) and how they appear as macro-behaviors when measured by statistical instruments (e.g., racially segregated urban areas) (Schelling, 1978).[2]

What the Appearance-from-Reality criterion demands in cases such as these is for our scientific theories to "save the phenomenon" not only in the sense that they should "deduce" or "predict" the appearances, but also in the sense that they should "produc[e]" them (van Fraassen, 2008, p. 283). Mere prediction, in other words, is not adequate for meeting this criterion; the theory should also provide the *mechanisms* that provide a bridge between the phenomenon and the appearances. van Fraassen shows that, while the Common Cause Principle is satisfied by the causal models of general use in natural and social sciences, the Appearance-from-Reality criterion is already rejected in fields such as cognitive science and quantum mechanics. The introduction of Big Data, it seems, is a development along the same trajectory, posing a threat to the Common Cause Principle as well as the Appearance-from-Reality criterion, repeating a historical pattern that has been the hallmark of scientific change for centuries.

**Saving the Phenomena: Causal Relations or Statistical Correlations?**

The distinction between relation and correlation is at the center of current debates among the proponents and detractors of "data-intensive science"—a debate that has been going on for decades, albeit with less intensity, between the advocates of *data-driven science* and those of *theory-driven science* (Hey, Tansley, & Toll, 2009). The proponents, who claim that "correlation supersedes causation, and science can advance even without coherent models, unified theories, or really any mechanistic explanation at all" (Anderson, 2008, para. 19), are, in some manner, questioning the viability of the *Common Cause Principle*, and calling for an end to theory, method, and the "old ways." Moving away from the "old ways", though, had begun even before the advent of Big Data, in the era of post-World War II Big Science when, for instance, many scientists involved in geological, biological, and ecological data initiatives held the belief that data collection could be an end in itself, independent of theories or models (Aronova, Baker, & Oreskes, 2010).

---

[2] We note that the relationship between the phenomenon and the appearance is starkly different in these two examples, with one having to do with (visual) perspective and the other with issues of emergence, reduction, and levels of analysis. Despite the difference, however, both examples speak to the gap between phenomenon and appearance, which is our interest here.

The defenders of the "old ways," on the other hand, respond in various manners. Some, such as microbiologist Carl Woese, sound the alarm for a "lack of vision" in an engineering biology that "might still show us how to get there; [but that] just doesn't know where 'there' is" (Woese, 2004, p. 173; cf. Callebaut, 2012, p. 71). Others, concerned about the ahistorical character of digital research, call for a return to the "sociological imagination" to help us understand what techniques are most meaningful, what information is lost, what data is accessed, etc. (Uprichard, 2012, p. 124). Still others suggest a historical outlook, advocating a focus on "long data" — that is, on datasets with a "massive historical sweep" (Arbesman, 2013, para. 4). Defining the perceived difference between the amount and nature of present and past data-driven sciences as superficially relying on sheer size, these authors suggest that one should focus on the dynamics of data accumulation (Strasser, 2012). Although data was historically unavailable on a petabyte scale, these authors argue, the problem of data overload existed even during the time when naturalists kept large volumes of unexamined specimens in boxes (ibid) and when social scientists were amassing large volumes of survey data (boyd & Crawford, 2012, p. 663).

A more conciliatory view comes from those who cast doubts on "using blind correlations to let data speak" (Harkin, 2013, para. 7) or from the advocates of "scientific perspectivism," according to whom "science cannot as a matter of principle transcend our human perspective" (Callebaut, 2012, p. 69; cf. Giere, 2006). By emphasizing the limited and biased character of *all* scientific representations, this view reminds us of the finiteness of our knowledge, and warns against the rationalist illusion of a God's-eye view of the world. In its place, scientific perspectivism recommends a more modest and realistic view of "science as engineering practice" that can also help us understand the world (Callebaut, 2012, p. 80; cf. Giere, 2006). Whether or not data-intensive science, enriched and promulgated by Big Data, is such a science is a contentious claim. What is perhaps not contentious is that science and engineering have traditionally been driven by distinctive principles. Based on the principle of "understanding by building," engineering is driven by the ethos of "working systems"—that is, those systems that function according to pre-determined specifications (Agre, 1997). While this principle might or might not lead to effectively working assemblages, its validity is certainly not guaranteed in the reverse direction: not all systems that work necessarily provide accurate theories and representations.  In other words, they save the appearance but not necessarily the phenomenon. This is perhaps the central dilemma of the twentieth century science, best manifested in areas such as artificial Intelligence and cognitive science (Ekbia, 2008). Big Data pushes this dilemma even further in practice and perspective, attached as it is to "prediction." Letting go of mechanical explanations, some radical perspectives on Big Data seek to save the phenomenon by simply saving the appearance. In so doing, they collapse the distinction between the two: phenomenon becomes appearance.

**Saving the Appearance: Productions or Predictions?**

Prediction is the hallmark of Big Data. Big Data allows practitioners, researchers, policy analysts, and others to predict the onset of trends far earlier than was previously possible. This ranges from the spread of opinions, viruses, crimes, and riots to shifting consumer tastes, political trends, and the effectiveness of novel medications and treatments. Many such social, biological, and cultural phenomena are now the object of modeling and simulation using the techniques of statistical physics (Castellano, Fortunato, & Loerto, 2009). British researchers, for instance, recently unveiled a program called Emotive "to map the mood of the nation" using data from Twitter and other social media (BBC, 2013). By analyzing two thousand tweets per second and rating them for expressions of one of eight human emotions (anger, disgust, fear, happiness, sadness, surprise, shame, and confusion), the researchers claim that they can "help calm civil unrest and identify early threats to public safety" (BBC, 2013, para. 3).

While the value of a project such as this may be clear from the perspective of law enforcement, it is not difficult to imagine scenarios where this kind of threat identification might function erratically, leading to more chaos than order. Even if we assume that human emotions can be meaningfully reduced to eight basic categories (what of complex emotions such as grief, annoyance, contentment, etc.?), and also assume cross-contextual consistency in the expression of emotions (how does one differentiate the "happiness" of the fans of Manchester United after a winning game from the expression of the "same" emotion by the admirers of the Royal family in the occasion of the birth of the heir to the throne?), technical limitations are quite likely to undermine the predictive value of the proposed system. Available data from Twitter have been shown to have a very limited character in terms of scale (spatial and temporal), scope, and quality (boyd & Crawford, 2012), making some observers wonder if this kind of "nowcasting" could blind us to the more hidden long-term undercurrents of social change (Khan, 2012).

In brief, a historical trend of which Big Data is a key component seems to have propelled a shift in terms of success criterion in science from causal explanations to predictive modeling and simulation. This shift, which is largely driven by forces that operate outside science (see the Political Economy section below), pushes the earlier trend—already underway in the 20th century, in breaking the bridge between phenomena and appearances—to its logical limit. If 19th century science strove to "save the phenomenon" and *produce* the appearance from it through causal mechanisms, and 20th century science sought to "save the appearance" and let go of causal explanations, 21st century science, and Big Data in particular, seems to be content with the *prediction* of appearances alone. The interest in the prediction of appearances, and the garnering of evidence needed to support them, still stems from the rather narrow deductive-nomological model, which presumes a law-governed reality. In the social sciences, this model may be rejected in favor of other explanatory forms that allow for the effects of human intentionality and free will, and that require the interpretation of the meanings of events and human behavior in a context-sensitive manner (Furner, 2004). Such alternative explanatory forms find little space in Big Data methodologies, which tend to collapse the distinction between phenomena and appearances altogether, presenting us with a structuralism run amok. As in the previous eras in history, this trend is going to have its proponents and detractors, both spread over a broad spectrum of views. There is no need to ask which of these embodies "real" science, for each one of them bears the seeds of its own dilemmas.

**METHODOLOGICAL DILEMMAS**

The epistemological dilemmas discussed in the previous section find a parallel articulation in matters of methodology in the sciences and humanities—e.g., in debates between the proponents of "qualitative" and "quantitative" methods. Some humanists and social scientists have expressed concern that "playing with data [might serve as a] gateway drug that leads to more-serious involvement in quantitative research" (Nunberg, 2010, para. 12), leading to a kind of technological determinism that "brings with it a potential negative impact upon qualitative forms of research, with digitization projects optimized for speed rather than quality, and many existing resources neglected in the race to digitize ever-increasing numbers of texts" (Gooding, Warwick, & Terras, 2012, para. 6-7). The proponents, on the other hand, drawing a historical parallel with "the domestication of human mind that took place with pen and paper," argue that the shift doesn't have to be dehumanizing: "Rather than a method of thinking with eyes and hand, we would have a method of thinking with eyes and screen" (Berry, 2011).

In response to the emerging trend of digitizing, quantifying, and distant reading of texts and documents in humanities, the skeptics find the methods of "data-diggers" and "stat-happy quants [that] take the human out of the humanities" (Parry, 2010, para. 5) and that "necessitate an impersonal invisible hand" (Trumpener, 2009, p. 164) at odds with principles of humanistic inquiry (Schöch, 2013, par. 7). According to these commentators, while humanistic inquiry has been traditionally epitomized by lone scholar[s], work in the digital humanities requires the participation of programmers, interface experts, and others (Terras, 2013, as quoted by Sunyer, 2013).

What might be in fact happening, though, is that research on Big Data calls into question the simplistic dichotomy between quantitative and qualitative methods. Even "fractal distinctions" (Abbott, 2001, p. 11) might not, as such, be adequate to describe the blurring that occurs between analysis and interpretation in Big Data, particularly given the rise of large-scale visualizations. Many of the methodological issues that arise, therefore, derive from the stage at which subjectivity is introduced into the process: that is, the decisions made in terms of sampling, cleaning, and statistical analysis. Proponents of a "pure" quantitative approach herald the autonomy of the data from subjectivity: "the data speaks for itself!" However, human intervention at each step puts into question the notion of a purely objective Big Data science.

**Data Cleaning: To Count or Not to Count?**
Revealing a historical pattern similar to what we saw in the sciences, these disputes echo long-standing debates in academe between qualitatively- and quantitatively-oriented practitioners: "It would be nice if all of the data which sociologists require could be enumerated," William Cameron wrote in 1963, "Because then we could run them through IBM machines and draw charts as the economists do…[but] not everything that can be counted counts, and not everything that counts can be counted" (p. 13). These words have a powerful resonance to those who observe the intermingling of raw numbers, mechanized filtering, and human judgment in the flow of Big Data.

Data making involves multiple social agents with potentially diverse interests. In addition, the processes of data generation remain opaque and under-documented (Helles & Jensen, 2013). As rich ethnographic and historical accounts of data and science in the making demonstrate, the data that emerges from such processes can be incomplete or skewed. Anything from instrument calibration and standards that guide the installation, development and alignment of infrastructures (Edwards, 2010) to human habitual practices (Ribes & Jackson, 2013) or even Intentional distortions (Bergstein, 2013) can disrupt the "rawness" of data. As Gitelman and Jackson (2013) point out, data need to be imagined and enunciated to exist as data, and such imagination happens in the particulars of disciplinary orientations, methodologies, and evolving practices.

Having been variously "pre-cooked" at the stages of collection, management and storage, Big Data does not arrive in the hands of analysts ready for analysis. Rather, in order to be "usable" (Loukides, 2012), it has to be "cleaned" or "conditioned" with tools such as Beautiful Soup and scripting languages such as Perl and Python (O'Reilly Media, 2011, p. 6). This essentially consists of deciding which attributes and variables to keep and which ones to ignore (Bollier, 2010, p. 13)—a process that involves mechanized human work, through services such as Mechanical Turk, but also interpretative human judgment and subjective opinions that can "spoil the data" (Andersen, 2010, c.f. Bollier, 2010, p. 13). These issues become doubly critical when personal data are involved on a large scale and "de-identification" turns into a key concern—issues that are further exacerbated by the potential for "re-identification," which, in turn, undermines the thrust in Big Data research toward 'data liquidity.' The dilemma between the ethos of data sharing, liquidity, and transparency, on one hand, and risks to privacy and anonymity through reidentification, on the other, emerges in such diverse areas as medicine, location-tagged payments, geo-locating mobile devices, and social media (Ohm, 2010; Tucker, 2013). Editors and publishing houses are increasingly aware of these tensions, particularly as organizations begin to mandate data sharing with reviewers and, following publication, with readers (Cronin, 2013).

The traditional tension between "qualitative" and "quantitative," therefore, is rendered obsolete with the introduction of Big Data techniques. In its place, we see a tension between the empirics of raw numbers, the algorithmics of mechanical filtering, and the dictates of subjective judgment, playing itself out in the question that Cameron raised more than fifty years ago: *what counts and what doesn't count*?

**Statistical Significance: To Select or Not to Select**

The question of what to include — what to count and what not to count — has to do with the "input" of Big Data flows. A similar question arises in regards to the "output," so to speak, in the form of a longstanding debate over the value of significance tests (e.g., Gelman & Stern, 2006; Gliner, Leech, & Morgan, 2002; Goodman, 1999; Goodman, 2008; Nieuwenhuis, Forstmann, & Wagenmakers, 2011). Traditionally, significance has been an issue with small samples because of concerns over volatility, normality, and validation. In such samples, a change in a single unit can reverse the significance and affect the reproducibility of results (Button et al., 2013). Furthermore, small samples preclude tests of normality (and other tests of assumptions), thereby invalidating the statistic (Fleishman, 2012). This might suggest that Big Data is immune from such concerns. Far from it—the large size of data, some argue, makes it perhaps *too* easy to find statistically significant results:

> As long as studies are conducted as fishing expeditions, with a willingness to look hard for patterns and report any comparisons that happen to be statistically significant, we will see lots of dramatic claims based on data patterns that don't represent anything real in the general population (Gelman, 2013b, para. 11).

boyd and Crawford (2012) call this phenomenon apophenia: "seeing patterns where none actually exist, simply because enormous quantities of data can offer connections that radiate in all directions" (p. 668). Similar concerns over "data dredging" and "cherry picking" have been raised by others (Taleb, 2013; Weingart, 2012, para, 13-15).

Various remedies have been proposed for improving false reliance on statistical significance, including the use of smaller p-values,[3] correcting for "researcher degrees of freedom," (Simmons, Nelson, & Simonsohn, 2011, p. 1359), and the use of Bayesian analyses (Irizarry, 2013). Other proposals, skeptical of the applicability to Big Data of "the tools offered by traditional statistics," suggest that we avoid using p-values altogether and rely on statistics that are "[m]ore intuitive, more compatible with the way the human brain perceives correlations"—e.g., rank correlations as potential models to account for the noise, outliers, and varying-sized "buckets" into which Big Data is often binned (Granville, 2013, para. 2-3; see also Gelman 2013a).

---
[3] That is, the probability of obtaining a test statistic at least as extreme as the observation, assuming a true null hypothesis.

These remedies, however, do not address concerns related to sampling and selection bias in Big Data research. Twitter, for instance, which has issues of scale and scope, is also plagued with issues of *replicability* in sampling. The largest sampling frame on Twitter, the "firehose," contains all public tweets but excludes private and protected ones. Other levels of access—the "gardenhose" (10% of public tweets), the "spritzer" (1% of public tweets), and "white-listed" (recent tweets retrieved through the API) (boyd & Crawford, 2012)—provide a high degree of variability, "randomness," and specificity, making the data prone to sampling and selection biases and replication nigh impossible. These biases are accentuated when people try to generalize beyond tweets or tweeters to all people within a target population, as we saw in the case of the Emotive project; as boyd and Crawford (2012) observed, "it is an error to assume 'people' and 'Twitter users' are synonymous" (p. 669).

In brief, Big Data research has not eliminated some of the major and longstanding dilemmas in the history of sciences and humanities; rather it has redefined and amplified them. Significant among these are the questions of *relevance* (what counts and what doesn't), *validity* (how meaningful the findings are), *generalizability* (how far the findings reach), and *replicability* (can the results be reproduced?). This kind of research also urges scholars to think beyond the traditional "quantitative" and "qualitative" dichotomy, challenging them to rethink and redraw their intellectual orientations along new dimensions.

## AESTHETIC DILEMMAS

If the character of knowledge is being transformed, traditional methods are being challenged, and intellectual boundaries redrawn by Big Data, perhaps our ways of *representing* knowledge are also re-imagined—and they are. Historians of science have carefully shown how scientific ideas, theories, and models have historically developed along with the techniques and technologies of imaging and visualization (Jones & Galison, 1998). They have also noted how "beautiful" theories in science tend to be borne out as accurate, even when they initially contradict the received wisdom of the day (Chandrasekhar, 1987). Bringing to bear the powerful visualization capabilities of modern computers, novel features of the digital medium, and their attendant dazzling aesthetics, Big Data marks a new stage in the relationship between "beauty and truth [as] cognate qualities" (Kostelnick, 2007, p. 283). Given the opaque character of large datasets, graphical representations become essential components of modeling and communication (O'Reilly Media, 2011). While this is often considered to be a straightforward "mapping" of data points to visualizations, a great deal of *translational* work is involved, which renders the accuracy of the claims problematic.

Data visualization traces its roots back to William Playfair (1759-1823) (Spence, 2001; Tufte, 1983), whose ideas "functioned as transparent tools that helped to make visible and understand otherwise hidden processes" (Hohl, 2011, p. 1039). Similar ideas seem to drive current interest in Big Data visualization, where "well-designed visual representations can replace cognitive calculations with simple perceptual inferences and improve comprehension, memory, and decision making" (Heer, Bostock & Ogievetsky, 2010, p. 59). Hohl (2011), however, contends that there is a fundamental difference between Playfair's approach and the current state of the art in Big Data. Rather than making hidden processes overt, he charges, "the visualisation [*sic.*] process has become hidden within a blackbox of hardware and software which only experts can fully comprehend" (Hohl, 2011, p. 1040). This, he says, flouts the basic principle of transparency in scientific investigation and reporting: we cannot truly evaluate the accuracy or significance of the results if those results were derived using nontransparent, aesthetically-driven algorithmic operations. While visualization has been heralded as a primary tool for clarity, accuracy, and sensemaking in a world of huge datasets (Kostelnick, 2007), visualizations also bring a host of additional, sometimes unacknowledged, tensions and tradeoffs into the picture.

**Data Mapping: Arbitrary or Transparent?**

The first issue is that, in order for a visualization to be rendered from a dataset, those data must be *translated* into some visual form—i.e., what is called "the principled mapping of data variables to visual features such as position, size, shape, and color" (Heer, Bostock, & Ogievetsky, 2010, p. 67; cf. Börner, 2007; Chen, 2006; Ware, 2013). By exploiting the human ability to organize objects in space, mapping helps users understand data via spatial metaphors (Vande Moere, 2005). Data, however, takes different forms, and this kind of translation requires a substantial conceptual leap that leaves its mark upon the resulting product. Considered from this perspective, then, "any data visualization is first and foremost a visualization of the conversion rules themselves, and only secondarily a visualization of the raw data" (Galloway, 2011, p. 88).

The spatial metaphor of mapping, furthermore, introduces some of the same limitations and tensions known to be intrinsic to any cartographic rendering of real-world phenomena. Not only must maps "lie" about some things in order to represent the truth about others, unbeknownst to most users any given map "is but one of an indefinitely large number of maps that might be produced … from the same data" (Monmonier, 1996, p. 2). These observations hold doubly true for Big Data, where the process of creating a visualization requires that a number of rather arbitrary decisions be made. Indeed, "for any given dataset the number of visual encodings—and thus the space of possible visualization designs—is extremely large" (Heer, Bostock, & Ogievetsky, 2010, p. 59). Despite technical sophistication, decisions about these visual encodings are not always carried out in an optimal way (Kostelnick, 2007, p. 285). Moreover, the generative nature of visualizations as a type of inquiry is often hidden— that is, some of the decisions are dismissed later as non-analytical and not worth documenting and reporting, while in fact they may have contributed to the ways in which the data will be seen and interpreted (Markham, 2013).

On a more foundational level, one could argue that 'mapping' may not even be the best label for what happens in data visualization. As with (bibliometric) 'maps' of science, the use of spatial metaphors and representations is not based on any epistemic logic, nor is it even logically cohesive because "science is a nominal entity, not a real entity like land masses are" (Day, forthcoming). Although a map is generally understood to be different from the territory that it represents, people also have an understanding of what features are relevant on a map. In the case of data visualization, however, the relevance of features is not self-evident, making the limits of the map metaphor and the attributions of meaning contestable. The actual layout of such visualizations, as an artifact of both the algorithm and physical limitations such as plotter paper, are in fact more frequently arbitrary than not (Ware, 2013). This, again, questions the relationship between phenomenon and appearance in Big Data analysis.

**Data Visualization: Accuracy or Aesthetics?**

An issue arises in the tension between accuracy of visualizations and their aesthetic appeal. As a matter of general practice, the rules employed to create visualizations are often weighted towards what yields a more visually pleasing result rather than directly mapping the data. Most graph drawing algorithms are based on a common set of criteria about what things to apply and what things to avoid (e.g. minimize edge crossing and favor symmetry), and such criteria strongly shape the final appearance of the visualization (Chen, 2006; Purchase, 2002). In line with the cognition-oriented perspective, proponents of aesthetically-driven visualizations point out that a more attractive display is likely to be more engaging for the end user (see Kostelnick, 2007; Lau & Vande Moere, 2007), maintaining that, while it may take users longer to understand the data in this format, they may "enjoy this (longer) time span more, learn complex insights … retain information longer or like to use the application repeatedly" (Vande Moere, 2005, pp. 172-173). Some have noted, however, that there is an inversely proportional relationship between the aesthetic and informational qualities of an information visualization (Gaviria, 2008; Lau & Vande Moere, 2007). As Galloway (2011) puts it, "the triumph of the aesthetic precipitates a decline in informatic perspicuity" (p. 98). Although the tension or tradeoff between a visualization's informational accuracy and its aesthetic appeal is readily acknowledged by scholars, these tradeoffs are not necessarily self-evident and are not always disclosed to users of data visualizations.

Even if such limitations were disclosed to the user, the relative role of aesthetic intervention and informational acuity is impossible to determine. Lau and Vande Moere (2007) have developed a model that attempts to classify any given visualization based on two continual, intersecting axes, each of which moves from accuracy to aestheticism (see Figure 2 for a rendering of this visualization).

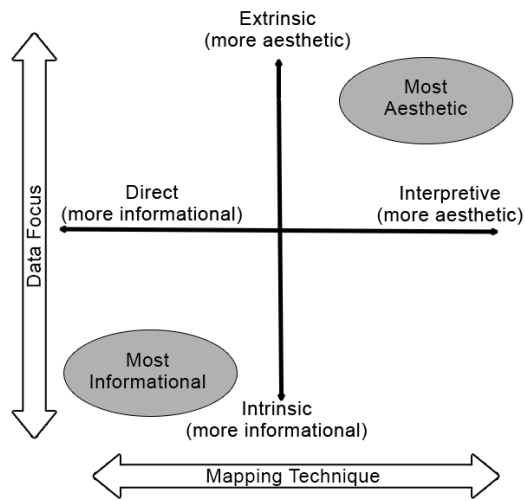

Figure 2. Rendering of Lau and Vande Moere's aesthetic-information matrix

The horizontal axis moves from literal representations of the numerical data (more accurate) to *interpretation* of the data (more aesthetic). The vertical axis represents, on one end, the degree to which the visualization is meant to help the user understand the data itself (more accurate), as compared to the opposite end, where the graphic is meant to communicate the *meaning* of the data (more aesthetic). Laying aside the larger problem of what constitutes an *interpretation* or *meaning* of a dataset, it is unclear how one could place a given visualization along either of these axes, as there is no clear rubric or metric for deciding the degree to which interpretations and images are reflective of meaning rather than data points. This makes the model hard to apply practically, and gets us no closer to the aim of providing useful information to the end user, who is likely unaware of the concessions that have been made so as to render a visualization aesthetically more pleasing and simultaneously less accurate. The model, however, brings out the tension that is generally at work in the visualization of data.

In summary, Big Data brings to fore the conceptual and practical dilemmas of science and philosophy that have to do with the relationship between truth and beauty. While the use of visual props in communicating idea and meanings is not new, the complexities associated with understanding Big Data and the visual capabilities of the digital medium have pushed this kind of practice to a whole new level, heavily tilting the balance between truth and beauty toward the latter.

**TECHNOLOGICAL DILEMMAS**

The impressive visual capabilities and aesthetic appeal of current digital products are largely due to the computational power of the underlying machinery. This power is provided by the improved performance of computational components, such as central processing units (CPUs), disk drive capacities, and input/output (I/O) speed and network bandwidth. Big Data storage and processing technologies generally need to handle very large amounts of data, and be flexible enough to keep up with the speed of I/O operations and with the quantitative growth of data (Adshead, n.d.). Real or near-real time information processing—that is, delivery of information and analysis results as they occur—is the goal and a defining characteristic of Big Data analytics. How this goal is accomplished depends on how computer architects, hardware designers, and software developers deal with the tension between continuity and innovation—a tension that drives technological development in general.

**Computers and Systems: Continuity or Innovation?**

The continuity of Big Data technologies is exemplified by such concepts as parallel, distributed, or grid computing, which date back to the 1950s and to IBM's efforts to carry out simultaneous computations on several processors (Wilson, 1994). The origins of cluster computing and high-performance computing (supercomputing), two groups of core technologies and architectures that maximize storage, networking, and computing power for data-intensive applications, can also be traced back to the 1960s. High-performance computing significantly progressed in the 21 century, pushing the frontiers of computing to petascale and exascale levels. This progress was largely enabled by software innovations in cluster computer systems, which consist of many "nodes," each having their own processors and disks, all connected by high-speed networks. Compared with traditional high-performance computing, which relies on efficient and powerful hardware, cluster computers maximize reliability and efficiency by using existing commodity (off-the-shelf) hardware and newer, more sophisticated software (Bryant, Katz, & Lasowska,

2008). A *Beowulf* cluster, for example, is a cluster of commodity computers that are joined into a small network, and use programs that allow processing to be shared among them. While software solutions such as Beowulf clusters may work well, their underlying hardware base limitations can seriously limit data-processing capacities as data continue to accumulate.

These limitations of software-based approaches call for hardware or "cyberinfrastructure" solutions that are met either by supercomputers *or* commodity computers. The choice between the two involves many parameters, including cost, performance needs, types of application, institutional support, and so forth. Larger organizations, including major corporations, universities, and government agencies, can choose to purchase their own supercomputers or buy time within a shared network of supercomputers, such as the National Science Foundation's XSEDE program.[4] With a two-week wait time and proper justification, anyone from a US university can get computing time on *Stampede*, the sixth fastest computer in the world (Lockwood, 2013). Smaller organizations with budget and support limitations would probably adopt a different approach that relies on commodity hardware and clusters.

The gap between these two approaches has created a space for big players that can invest significant resources in innovative and customized solutions suited specifically to deal with Big Data, with a focus on centralization and integration of data storage and analytics (see the discussion on computerization movements above).[5] This gave rise to the concept of the "data warehouse" (Inmon, 2000)—a predecessor of "cloud computing" that seeks to address the challenges of creating and maintaining a centralized repository. Many data warehouses run on software suites that manage security, integration, exploration, reporting, and other processes necessary for big data storage and analytics, including specialized commercial software suites such as EMC Greenplum Database Software and open source alternatives such as Apache Hadoop-based solutions (Dumbill, 2012). These solutions, in brief, have given rise to a spectrum of alternative solutions to the continuity-innovation dilemma. Placed on this spectrum, Hadoop can be considered a compromise solution; it is designed to work on commodity servers and yet it stores and indexes both structured and unstructured data (Turner, 2011). Many commercial enterprises such as Facebook utilize the Hadoop framework and other Apache products (Thusoo et al., 2010). Hive is an addition that deals with the continuity-innovation dilemma by bringing together the "best" of older and newer technologies, making use of traditional data warehousing tools such as SQL, metadata, and partitioning to increase developers' productivity and stimulate more collaborative ad hoc analytics (Thusoo et al., 2010). The popularity of Hadoop has as much to do with performance as it does with the fact that it runs on commodity hardware. This provides a clear competitive advantage to large players such as Google that can rely on freely available software and yet afford significant investments in purchasing and customizing hardware and software components (Fox, 2010; Metz, 2009). This would also reintroduce the invisible divide between "haves"—those who can benefit from the ideological drive toward "more" (more data, more storage, more speed, more capacity, more results, and, ultimately, more profit or reward)—and "have nots" who lag behind or have to catch up by utilizing other means.

---

[4] https://www.xsede.org/
[5] The largest data warehousing companies include eBay (5 petabytes of data), Wal-Mart Stores, (2.5 petabytes), Bank of America (1.5 petabytes), and Dell (1 petabyte) (Lai, 2008). Amazon.com operated a warehouse in 2005 consisting of 28 servers and a storage capacity of over 50 terabytes (Layton, 2005).

**The Role of Humans: Automation or Heteromation?**

Whatever approach is adopted in dealing with the tension between continuity and innovation, the technologies behind Big Data projects have another aspect that differentiates them from earlier technologies of automation, namely the changing role and scale of human involvement.

While earlier technologies may have been designed with the explicit purpose of replacing human labor or minimizing human involvement in socio-technical systems, Big Data technologies heavily rely on human labor and expertise in order to function. This shift in the technological landscape drives the recent interest in crowdsourcing and in what De Roure (2012) calls "the rise of social machines." As more computers and computing systems are connected into cyberinfrastructures that support Big Data, more and more citizens participate in the digital world and form a Big Society that helps to create, analyze, share, and store Big Data. De Roure mentions citizen science and volunteer computing as classic examples of a human-machine symbiosis that transcends purely technical or computational solutions.

Another aspect of current computing environments that calls for further human involvement in Big Data is the changing volume and complexity of software code. As software is getting more and more sophisticated, code defects, such as bugs or logical errors, are getting harder and harder to identify. Testing that is done in a traditional way may be insufficient regardless of how thorough and extensive it is. Increasingly aware of this, some analysts propose opening up all software code, sharing it with others along with data and using collaborative testing environments to reproduce data analysis and strengthen credibility of computational research (Stodden, Hurlin, & Perignon, 2012; Tenner, 2012).

Crowdsourcing or social machines leverage the labor and expertise of millions of individuals in the collection, testing or otherwise processing of data, carrying out tasks that are beyond current computer capabilities, and solving formidable computational problems. For example, NASA's Tournament Lab challenged its large distributed community of researchers to build software that would automatically detect craters on various planets (Sturn, 2011). The data used by developers to create their algorithms were created using Moon Zoo, a citizen science project at zooniverse.com that studies lunar surface. Crater images manually labeled by thousands of individuals were used for creating and testing the software. The FoldIt project is another citizen science project that utilizes a gaming approach to unravel very complex molecular structures of the proteins.[6] A similar process of dividing large tasks into smaller ones and using many people to work on them in parallel is used by Amazon's Mechanical Turk Project, although it involves some financial compensation. Amazon advertises its approach as "access to an on-demand, scalable workforce."[7]

The shared tenet among these various projects is their critical reliance on technologies that require active human involvement on large scales. Rather than describing the use of machines in Big Data in terms of automation, perhaps we should acknowledge the continuing creative role of humans in knowledge infrastructure and call it "heteromation" (Ekbia & Nardi, forthcoming). Heteromation and the rise of social machines also highlights the distinction between data generated *by* people versus data generated *about* people. This kind of "participatory personal data" (Shilton, 2012) describes a new set of practices where individuals contribute to data collection by using mediating technologies, ranging from web entry forms to GPS trackers and sensors to more sophisticated apps and games. These practices shift the dynamics of power in the acts of measurement, groupings and classifications between the measuring and the measured subjects (Nafus, Sherman, under review). Humans who have always been the measured subjects receive an opportunity to "own" their data, challenge the metrics or even expose values and biases embedded in automated classifications and divisions (Dwork and Mulligan, 2013). Whether or not this kind of "soft resistance" will tilt the balance of power in favor of small players is yet to be seen, however.

---

[6] http://fold.it/portal/info/science
[7] https://www.mturk.com/mturk/

In brief, technology has evolved in concert with the Big Data movement, arguably providing the impetus behind it. These developments, however, are not without their concerns. The competitive advantage for institutions and corporations that can afford the computing infrastructure necessary to analyze Big Data creates a new subset of technologically elite players. Simultaneously, there is a rise in shared "participation" in the construction of Big Data through crowdsourcing and other initiatives. The degree to which these developments represent opportunities or undermine ownership, privacy, and security remains to be discussed.

## LEGAL AND ETHICAL DILEMMAS

Changes in technology, along with societal expectations that are partly shaped by these changes, raise significant legal and ethical problems. These include new questions about the scope of individual privacy and the proper role of intellectual property protection. Inconveniently, the law affords no single principle to balance the competing interests of individuals, industries, and society as a whole in the burgeoning age of Big Data. As a result, policymakers must negotiate a new and shifting landscape (Solove, 2013, p. 1890).

### Privacy Concerns: To Participate or Not?

The concept of privacy as a generalized legal "right" was introduced in an 1890 Harvard Law Review article written by Samuel Warren and Louis Brandeis—both of whom would later serve as United States Supreme Court Justices (Warren & Brandeis, 1890). "The right to be let alone," the authors argued, is an "inevitable" extension of age-old legal doctrines that discourage intrusions upon the human psyche (p. 195). These doctrines include laws forbidding slander, libel, nuisances, assaults, and the theft of trade secrets (pp. 194, 205, 212). Since the time of Warren and Brandeis, new technologies have led policymakers to continually reconsider the definition of privacy. In the 1980s and 1990s, the widespread adoption of personal computers and the Internet led to the enactment of statutes and regulations that governed the privacy of electronic communications (Bambauer, 2012, p. 232). In the 2000s, even more laws were enacted to address *inter alia*, financial privacy, healthcare privacy, and children's privacy (Schwartz & Solove, 2011, p. 1831). Today, Big Data presents a new set of privacy concerns in diverse areas such as health, government, intelligence, and consumer data.

One new privacy challenge stems from the use of Big Data to predict consumer behavior. A frequently (over)used example is that of the department store Target predicting the pregnancy of expectant mothers in order to target them with coupons at what is perceived to be the optimal time (Duhigg, 2012). In a similar vein, advertisers have recently started assembling shopping profiles of individuals based on compilations of publicly available metadata (such as the geographic locations of social media posts) (Mattioli, forthcoming), which has led some commentators to criticize such practices as privacy violations (Tene & Polonetsky, 2012). Similar concerns have been expressed in the realm of medical research, where the electronic storage and distribution of individual health data can potentially reveal information not only about the individual but about others related to them. This might happen, for instance, with the public availability of genomic data, as in a recent case that raised objections from the family members of a deceased woman. The subsequent removal of that information marked a legal triumph for the family, but was a worrisome sign for advocates of open access that privacy concerns might significantly slow the progress of Big Data research (Zimmer, 2013). As leading commentators have observed, "[t]hese types of harms do not necessarily fall within the conventional invasion of privacy boundaries," and thus raise new questions for policymakers (Crawford & Schultz, 2014).

Another set of privacy concerns stems from government uses of Big Data. The recent revelation in the U.S. about the National Security Agency (NSA)'s monitoring of email and mobile communications of millions of Americans might just be the tip of a much bigger legal iceberg that will affect the whole system, including the Supreme Court (Risen, 2013; Risen & Lichtblau, 2013). Although the public seems to be divided in its perception of the privacy violations (Shane, 2013), the official defense by the government tends to highlight the point that data gathering did not examine the substance of emails and phone calls, but rather focused on more general metadata (Savage & Shear, 2013). Similar themes were raised in a 2012 Supreme Court ruling on the constitutionality of the use of GPS devices by police officers to track criminal suspects (U.S. v. Jones, 2012). Public discussions on approaches to and standards of privacy are complicated by the fact that government officials seem to be concerned less with what they actually do to people's data than with public perceptions on what they do or might do (Nash, 2013).

The common theme that emerges from the foregoing examples is that vastly heterogeneous types of data can be generated, transferred, and analyzed without the knowledge of those affected. Such data are generated silently and often put to unforeseen uses after they have been collected by known and unknown others, implicating privacy but also leading to second-order harms, such as "profiling, tracking, discrimination, exclusion, government surveillance and loss of control" (Tene & Polonestsky, 2012, p. 63). Ironically, the protection of privacy, as well as its violation, depends on technology just as much as it depends on sound public policy. "Masking" that seeks to obfuscate personally identifying information while preserving the usefulness of underlying data, for instance, employs sophisticated encryption techniques (El Emam, 2011). Despite its sophistication, however, it has been shown to be vulnerable to re-identification, leading Ohm (2010) to opine that "[r]eidentification science disrupts the privacy policy landscape by undermining the faith that we have placed in anonymization" (p. 1704). Another technological solution to the privacy puzzle—namely, using software to meter and track the usage of individual parcels of data—requires individual citizens and consumers to tag their data with their privacy preferences. This approach, which was put forth by the World Economic Forum, portrays a future in which banks, governments, and service providers would supply consumers with personally identifiable data collected about them (World Economic Forum, 2013, p. 13). By resorting to (meta)data to protect data, though, this "solution" puts the onus of privacy protection on individual citizens. As such, it reintroduces in stark form the old dilemma of the division of labor and responsibility between the public and private spheres.

**Intellectual Property: To Open or to Hoard?**

The risk of free riding is a collective action problem well known to intellectual property theorists. Resources that are costly to produce and subject to cheap duplication tend to be underproduced because potential producers have little to gain and everything to lose. The function of intellectual property laws is to create an incentive for people to produce such resources by entitling them to enjoin copyists for a limited period of time. Conventional data, however, do not meet the eligibility requirements for patent protection, and are often barred from copyright protection because commercially published data are often factual in nature (Patent Act, Copyright Act)(When conventional data has met the eligibility requirements of copyright, protection has generally been thin.) This facet of American intellectual property law has led to efforts by database companies to urge Congress to enact new laws that would provide *sui generis* intellectual-property protection to databases (Reichman & Samuelson, 1997).

Big Data may mark a new chapter in the story of intellectual property, expanding it to the broader issue of *data ownership*. Ironically, the very methods and practices that make Big Data useful may also infuse it with subjective human judgments. As discussed earlier, a researcher's subjective judgments can become deeply infused into a data set through sampling, data cleaning, and creative manipulations such as data masking. As a result, Big Data compilations may actually be more likely to satisfy the prerequisites for copyrightability than canonical factual compilations (Mattioli, forthcoming). If so, *sui generis* database protection may be unnecessary. Existing intellectual property laws may also need to be adapted in order to accommodate Big Data practices. In the United Kingdom, for example, lawmakers have approved legislation that provides a copyright exemption for data mining, allowing search engines to copy books and films in order to make them searchable. U.S. lawmakers have yet to seriously consider such a change to the Copyright Act.

To recap, the ethical and legal challenges brought about by Big Data present deeper issues that suggest significant changes to dominant legal frameworks and practices. In addition to privacy and data ownership, Big Data challenges the conventional wisdom of collective action phenomena such as free riding—a topic discussed in the following section.

## POLITICAL ECONOMY DILEMMAS

The explosive growth of data in recent years is accompanied by a parallel development in the economy — namely, an explosive growth of wealth and capital in the global market. This creates an interesting question about another kind of correlation, which, unlike the ones that we have examined so far, seems to be extrinsic to Big Data. Crudely stated, the question has to do with the relationship between the growth of data and the growth of wealth and capital. What makes this a particularly significant but also paradoxical question is the concomitant rise in poverty, unemployment, and destitution for the majority of the global population. A distinctive feature of the recent economic recession is its strongly *polarized* character: a large amount of data and wealth is created, giving rise to philanthropic projects in the distribution of both data and money, concentrated in the hands of a very small group of people. This gives rise to a question regarding the relation between data and poverty: it can be argued that Big Data tends to *generally* act as a polarizing force not only in the market, but also in arenas such as science.

The exploration of these questions leads to the issue of the *mechanisms* that support and enable such a polarizing function. Variegate in nature, such mechanisms operate at the psychological, socio-cultural, and political levels, as described below.

**Data as Asset: Contribution or Exploitation?**

The flow of data on the Internet is largely concentrated in social networking sites, meta-search engines, gaming, and, to a lesser degree, in science (www.alexa.com/topsites). These same sites, it turns out, also represent a large proportion of the flow of capital in the so-called information economy. Facebook, for instance, increased its ad revenue from $300 million in 2008 to $4.27 billion in 2012. The observation of these trends reinforces the World Economic Forum's recognition of data as a new "asset class" (2011) and the notion of data as the "new oil". The caveat is that data, unlike oil, is not a natural resource, which means that its economic value cannot derive from what economists call "rent." What is the source of the value, then?

This question is at the center of an ongoing debate that involves commentators from a broad spectrum of social and political perspectives. Despite their differences, the views of many of these commentators converge on a single source: "users." According to the VP for Research of the technology consulting firm Gartner, Inc., "Facebook's nearly one billion users have become the largest unpaid workforce in history" (Laney, 2012). From December 2008 to December 2013, the number of users on Facebook went from 140 million to more than one billion. During this same period of time, Facebook's revenue rose by about 1300%. Facebook's filing with the Securities and Exchange Commission in February 2012 indicated that "the increase in ads delivered was driven primarily by user growth" (Facebook, 2012, p. 50). Turning the problem of free riding on its head, these developments introduce a novel phenomenon, where instead of costly resources being underproduced because they can be cheaply duplicated (see the discussion of data ownership above), user data generated at almost no cost are overproduced, giving rise to vast amounts of wealth concentrated in the hands of proprietors of technology platforms. The character of this phenomenon is the focus of the current debate.

Fuchs (2010), coming to the debate from a Marxist perspective, has argued that users of social media sites such as Facebook are exploited in the same fashion that TV spectators are exploited.[8] The source of exploitation, according to Fuchs, is the "free labor" that users put into the creation of user-generated content (Terranova, 1999). Furthermore, the fact that users are not financially compensated throws a very diverse group of people into an exploited class that Fuchs, following Hardt and Negri (2000), calls the "multitude."

The nature of user contribution finds a different characterization by Arvidsson and Colleoni (2012), who argue that the economy has shifted toward an affective law of value "where the values of companies and their intangible assets are set not in relation to an objective measurement, like labor time, but in relation to their ability to attract and aggregate various kinds of affective investments, like intersubjective judgments of their overall value or utility in terms of mediated forms of reputation" (p. 142). This leads these authors to the conclusion that the right explanation for the explosive wealth of companies such as Facebook should be sought in financial market mechanisms such as branding and valuation.

---

[8] The theory of audience exploitation was originally put forth by the media theorist Dallas Smythe (1981).

Conversely, Ekbia (forthcoming) contends that the nature of user contribution should be articulated in the digitally mediated networks that are prevalent in the current economy. The winners in this "connexionist world" (Boltanski & Ciapello, 2007) are the flexibly mobile, those who are able to move not only geographically (between places, projects, and political boundaries), but also socially (between people, communities, and organizations) and cognitively (between ideas, habits, and cultures). This group largely involves the *nouveau riche* of the Internet age (e.g., the founders of high-tech communications and social media companies) (Forbes, 2013). The "losers" are those who have to play as stand-ins for the first group in order for the links created in these networks to remain active, productive, and useful. Interactions between these two groups are embedded in a form of organizing that can be understood as "expectant organizing"—a kind of organization that is structured with *built-in* holes and gaps that are intended to be bridged and filled through the activities of end users.[9]

Whichever of the above views one considers as an explanation for the source of value of data, it is hard not to acknowledge a correlation between user participation and contribution and the simultaneous rise in wealth and poverty.

**Data and Social Image: Compliance or Resistance?**

Every society creates the image of an "idealized self" that presents itself as the archetype of success, prosperity, and good citizenship. In contemporary societies, the idealized self is someone who is highly independent, engaged, and self-reliant—a high-mobility person, as we saw, with a potential for re-education, re-skilling, and relocation; the kind of person often sought by cutting-edge industries such as finance, medicine, media, and high technology. This ''new man,'' according to sociologist Richard Sennett (2006), "takes pride in eschewing dependency, and reformers of the welfare state have taken that attitude as a model—everyone his or her own medical advisor and pension fund manager" (p. 101). Big Data has started to play a critical role in both propagating the image and developing the model, playing out a three-layer mechanism of social control through monitoring, mining, and manipulation. Individual behaviors, as we saw in the discussion of privacy, are under continuous monitoring through Big Data techniques. The data that are collected are then mined for various economic, political, and surveillance purposes: Corporations use the data for targeted advertising, politicians for targeted campaigns, and government agencies for targeted monitoring of all manners of social behavior (health, finance, criminal, security, etc.). What makes these practices particularly daunting and powerful is their capability in identifying patterns that are *not* detectable by human beings, and are indeed unavailable before they are mined (Chakrabarti, 2009). Furthermore, these same patterns are fed back to individuals through mechanisms such as recommendation systems, creating a vicious cycle of regeneration that puts people in "filter bubbles" (Pariser, 2012).

These properties of autonomy, opacity, and generativity of Big Data bring the game of social engineering to a whole new level, with its attendant benefits and pitfalls. This leaves the average person with an ambivalent sense of empowerment and emancipatory self-expression combined with anxiety and confusion. This is perhaps the biggest dilemma of contemporary life, which rarely disappears from the consciousness of the modern individual.

---

[9] To understand this, we need to think in terms of *networks* (in the plural) rather than *network* (in the singular). The power and privilege of the winners is largely in the fact that they can move across various networks, creating connections among them—a privilege that others are deprived of because they need to stay put in order to be included. This is the key idea behind the notion of "structural holes," although the standard accounts in the organization and management literature present it from a different perspective (e.g., Burt, 1992).

---

## DISCUSSION

We have reviewed here a large and diverse literature. We could, no doubt, have expanded upon each of these sections in book-length form as each section identifies a diversity of data and particularlistic issues associated with these data. However, the strength of this synthesis is in the identification of recurring issues and themes across these disparate domains. We would like to discuss these themes briefly here.

**Big Data is not a monolith**

The review reveals, as expected, the diversity and heterogeneity of perspectives among theorists and practitioners regarding the phenomenon of Big Data. This starts with the definitions and conceptualizations of the phenomenon, but it certainly does not stop there. Looking at Big Data as a product, a process, or a new phenomenon that challenges human cognitive capabilities, commentators arrive at various conclusions in regards to what the key issues are, what solutions are available, and where the focus should be. In dealing with the cognitive challenge, for instance, some commentators seek the solution by focusing on *what* to delete (Waldrop, 2008, p. 437), while others, noting the increasing storage capacity of computers at decreasing costs, suggest a different strategy: purchasing more storage space and keeping everything (Kraska, 2013, p. 84).

The perspective of computerization movement suggested here introduces a broader socio-historical perspective, and highlights the gap between articulated visions and the practical reality of Big Data. As in earlier major developments in the history of computing, this gap is going to resurface in the tensions, negotiations, and partial resolutions among various players, but it is not going to disappear by fiat, nor does it surface in precisely the same way across all domains (e.g., health, finance, science). Whether we deal with the tension between the humanities and administrative agendas in academia, or between business interests and privacy concerns in law, the gap needs to be dealt with in a mixture of policy innovation and practical compromise, but it cannot be wished away. The case of British law exemption to copyright in data mining provides a good example of such pragmatic compromise (Owens, 2011).

**The light and dark side of Big Data**

Common attributes of Big Data—not only the five v's of volume, variety, velocity, value, and veracity, but also its autonomy, opacity, and generativity—endow this phenomenon with a kind of novelty that is productive and empowering yet constraining and overbearing. Thanks to Big Data techniques and technologies, we can now make more accurate predictions and potentially better decisions in dealing with health epidemics, natural disasters, or social unrest. At the same time, however, one cannot fail to notice the imperious opacity of data-driven approaches to science, social policy, cultural development, financial forecasting, advertising, and marketing (Kallinikos, 2013). In confronting the proverbial Big Data elephant, it seems we are all blind, regardless of how technically equipped and sophisticated we might be. We saw the different aspects of the opacity in discussing the size of data and in the issues that arise when one tries to analyze, visualize, or monetize Big Data projects. *Big data is dark data*—this is a lesson that financial analysts and forecasters learned, and the rest of us paid for it handsomely, during the recent economic meltdown. The fervent enthusiasts of technological advance refuse to admit and acknowledge this side of the phenomenon, whether they are business consultants, policy advisors, media gurus, or scientists designing eye-catching visualization techniques. The right approach, in response to this unbridled enthusiasm, is not to deny the light side of Big Data, but rather to devise techniques that bring human judgment and technological prowess to bear in a meaningfully balanced manner.

**The Futurity of Big Data**

We live in a society that is obsessed with the future and predicting it. Enabled by the potentials of modern technoscience—from genetically-driven life sciences to computer-enabled social sciences—contemporary societies take more interest in what the future, as a set of possibilities, holds for us than how the past emerged from a set of possible alternatives; the past is interesting, if at all, only insofar as it teaches us something about the future. As Rose and Abi-Rached (2013) wrote, "we have moved from the risk management of almost everything to a general regimes of futurity. The future now presents us neither with ignorance nor with fate, but with probabilities, possibilities, a spectrum of uncertainties, and the potential for the unseen and the unexpected and the untoward" (p. 14).

Big Data is the product, but also an enabler, of this regime of futurity, which is embodied in Hal Varian's notion of "nowcasting" (Varian, 2012). The ways by which Big Data shapes and transforms various arenas of human activity vary depending on the character of the activity in question. In health, where genetic risks and behavioral dispositions have turned into foci of diagnosis and care, techniques of data mining and analytics provide very useful tools for the introduction of preventative measures that can mitigate future vulnerabilities of individuals and communities. Meteorologists, security and law enforcement agencies, and marketers and product developers can similarly benefit from the predictive capacity of these techniques. Applying the same predictive approach to other areas such as social forecasting, academic research, cultural policy, or inquiries in the humanities, however, might produce outcomes that are narrow in reach, scope, and perspective, as we saw in the discussion of Twitter earlier in this paper. With data having a lifetime on the order of minutes, hours, or days, quickly losing its meaning and value, caution needs to be exercised in making predictions or truth claims on matters with high social and political impact. In the humanities, for example, the discipline of history can be potentially revitalized through the application of modern computer technologies, allowing historians to ask new questions and to perhaps find new insights into the events of the near and far past (Ekbia & Suri, 2013). At the same time, we cannot trivialize the perils and pitfalls of a historical inquiry that derives its clues from computer simulations that churn massive but not necessary data (Fogu, 2009).

**The Disparity of Big Data**

Research and practice in Big Data require vast resources, investments, and infrastructure that are only available to a select group of players. Historically, technological developments of this magnitude have fallen within the purview of government funding, with the intention (if not the guarantee) of universal access to all. The most recent case of this magnitude was the launch of the Internet in the U.S., touted as one of the most significant contributions of the federal government to technological development on a global scale. Although this development has taken a convoluted path, giving rise to (among other things) what is often referred to as the "digital divide" (Buente & Robbin, 2008), the historical fact that this was originally a government project does not change. Big Data is taking a different path. This is perhaps the first time in modern history, particularly in the U.S., that a technological development of this magnitude has been left largely in the hands of the private sector. Initiatives such as the National Science Foundation's XSEDE program, discussed earlier, are attempt in alleviating the disparities caused by the creation of a technological elite. Although the full implications of this are yet to be revealed, disparities between the haves and have-nots are already visible between major technology companies and their users, between government and citizens, and between large and small businesses and universities, giving rise to what can be called the "data divide." This is not limited to divisions between sectors, but also creates further divides *within* sectors. For example, technologically-oriented disciplines will receive priority for funding and other resources given this computationally-intensive turn in research. This potentially leads to a Matthew Effect (Merton, 1973) reoriented around technological skills and expertise rather than publication and citation.

## BEYOND DILEMMAS: WAYS OF ACTING

Digital technologies have a dual character—empowering, liberating, and transparent, yet also intrusive, constraining, and opaque. Big Data, as the embodiment of the latest advances in digital technologies, manifests this dual character in a vivid manner, putting the tensions and frictions of modernity in high relief. Our framing of the social, technological, and intellectual issues arising from Big Data as a set of dilemmas was based on this observation. As we have shown here, these dilemmas are partly continuous with those of earlier eras but also partly novel in terms of their scope, scale, and complexity— hence, our metaphorical trope of "bigger dilemmas."

Dilemmas, of course, do not easily lend themselves to solutions or resolutions. Nonetheless, they should not be understood as impediments to action. In order to go beyond dilemmas, one needs to understand their historical and conceptual origins, the dynamics of their development, the drivers of the dynamics, and the alternatives that they present. This is the tack that we followed here. This allowed us to trace out the histories, drivers, and alternative pathways of change along the dimensions that we examined. Although these dimensions do not exhaustively cover all the socio-cultural, economic, and intellectual aspects of change brought about by Big Data, they were broad enough to provide a comprehensive picture of the phenomenon and to reveal some of its attributes that have been hitherto unnoticed or unexamined in a synthesized way. The duality, futurity, and disparity of Big Data, along with its various conceptualizations among practitioners, make it unlikely for a consensus view to emerge in terms of dealing with the dilemmas introduced here.  Here, as in science, a perspectivist approach might be our best bet. Such an approach is based on the assumption that "all theoretical claims remain perspectival in that they apply only to aspects of the world and then, in part because they apply only to some aspects of the world, never with complete precision" (Giere, 2006, p. 15). Starting with this assumption, and keeping the dilemmas in mind, we consider the following as potential areas of research and practice for next steps.

**Ways of Knowing**

Many of the epistemological critiques of Big Data have centered around the assertion that theory is no longer necessary in a data-driven environment. We must ask ourselves what it is about this claim that strikes scholars as problematic. The virtue of a theory, according to some philosophers of science, is its intellectual economy — that is, its ability to allow us to obtain and structure information about observables that are otherwise intractable (Duhem, 1906, 21ff). "Theorists are to replace reports of individual observations with experimental laws and devise higher level laws (the fewer, the better) from which experimental laws (the more, the better) can be mathematically derived" (Bogen, 2013). In light of this, one wonders if objections to data-driven science have to do with the practicality of such an undertaking (e.g., because of the bias and incompleteness of data), or there are things that theories could tell us that Big Data cannot? This calls for deeper analyses on the role of theory and the degree to which Big Data approaches challenge or undermine that role. What new conceptions of theory, and of science for that matter, do we need in order to accommodate the changes brought about by Big Data? What are the implications of these conceptions for theory-driven science? Does this shift change the relationship between science and society? What type of knowledge is potentially lost in such a shift?

**Ways of Being Social**

The vast quantity of data produced on social media sites, the ambiguity surrounding data ownership and privacy, and the nearly ubiquitous participation in these sites makes them an ideal locus for studies of Big Data. Many of the dilemmas discussed in this paper find a manifestation in these settings, turning them into useful sources of information for further investigation. One area of research that requires more attention is the manner in which these sites contribute to identity construction at the micro, meso, and macro levels. How is the image of the idealized self (Weber, 1904) enacted and how do people frame their identities (Goffman, 1974) in current digitized environments? What is the role of ubiquitous micro-validations in this process? To what extent do these platforms serve as mechanisms for liberating and in what ways do they serve to constrain behavior? When do projects such as Emotive become not only descriptive of the emotions in a society, but self-fulfilling prophecies that regenerate what they allegedly uncover?

**Ways of Being Protected**

The current environment seems to offer little in way of protecting personal rights, including the right to personal information. Issues of privacy are perhaps among the most pressing and timely issues, particularly in light of recent revelations about the access by National Security Agency in the U.S. and other governmental and non-governmental organizations around the globe. Regulation ensures citizens a "reasonable expectation of privacy." However, boundaries of what is expected and reasonable are blurred in the context of Big Data. New notions of privacy need to take into account the complexities (if not impossibility) of anonymity in the Big Data age and to negotiate between the responsibilities and expectations of producers, consumers, and disseminators of data. Specific contexts should be investigated in which privacy is most sensitive — for example, gene patents and genetic databases. It is hypothesized that such investigations would reveal the diversity of Big Data policies that will be necessary to take into account the differences, as well as relationships, among varying sectors. Issues of risk and responsibility are

paramount—what level of responsibility must citizens, corporations, and the government absorb in terms of protecting information, rights, and accountabilities? What reward and reimbursement mechanisms should be implemented for an equitable distribution of wealth created through user-generated content? Even more fundamentally, what alternative socio-economic arrangements are conceivable to reverse the current trend toward a heavily polarized economy?

**Ways of Being Technological**

The division of labor between machines and humans has been a recurring motif of modern discourse. The advent of computing has shifted the boundaries of this division in ways that are both encouraging and discouraging in regards to human prosperity, dignity, and freedom. While technologies do not develop on their own accord, they do have the capacity to impose a certain logic not only on how they relate to human beings but also on how human beings relate to each other. This brings up a host of questions that are novel or under-explored in character: What is the right balance between automation and heteromation in large socio-technical systems? What socio-technical arrangements of people, artifacts and connections make Big Data projects work and not work? What are the potentials and limits of crowdsourcing in helping us meet current challenges in science, policy, and governance? What types of infrastructures are needed to enable technological innovation, and who (the public or the private sector) should be their custodians? What parameters influence decisions in the choice of hardware and software? How can we study these processes? As data becomes big and heterogeneous, how do we store and describe it for long-term preservation and re-use? Can we do ethnographies of supercomputing and other Big Data infrastructures? And so forth.

In brief, the depth, diversity, and complexity of questions and issues facing our societies is proportional, if not exponential, to the amount of data flowing in the social and technological networks that we have created. More than five decades ago, Weinberg (1961) wrote with concern on the shift from "little science" to "big science" and the implications for science and society:

> Big Science needs great public support [and] thrives on publicity. The inevitable result is the injection of a journalistic flavor into Big Science, which is fundamentally in conflict with the scientific method. If the serious writings about Big Science were carefully separated from the journalistic writings, little harm would be done. But they are not so separated. Issues of scientific or technical merit tend to get argued in the popular, not the scientific, press, or in the congressional committee room rather than in the technical society lecture hall; the spectacular rather than the perceptive becomes the scientific standard. (p. 161).

Substitute Big Data for Big Science in these remarks, and one could easily see the parallels between the two situations, and the need for systematic investigation of current developments. What makes this situation different, however, is the degree by which all people from all walks of life are affected by ongoing developments. We would not, therefore, argue that conversations about Big Data should be removed from the popular press or the congressional committee room. On the contrary, the complexity and ubiquity of Big Data requires a concerted effort between academics and policy makers, in rich conversation with the public. We hope that this critical review provides the platform for the launch of such efforts.

**CITED REFERENCES**